\documentclass[twocolumn]{aastex62}

\usepackage{amsmath,amsbsy,bm,color,graphics,multirow,footmisc,hyperref,etoolbox}
\hypersetup{
  colorlinks = true,
  urlcolor   = blue,
  linkcolor  = blue,
  citecolor  = blue
}

\pdfoutput=1


\begin{document}

\title{Evidence for the Cross-correlation between Cosmic Microwave Background Polarization Lensing from \textsc{Polarbear}{} and Cosmic Shear from Subaru Hyper Suprime-Cam}

%\submitjournal{ApJ}
\correspondingauthor{Toshiya Namikawa, Yuji Chinone}
\email{namikawa.toshiya9@gmail.com, chinoney@berkeley.edu}

\author[0000-0003-3070-9240]{T. Namikawa}
\affiliation{Department of Applied Mathematics and Theoretical Physics, University of Cambridge, Cambridge CB3 0WA, UK}

\author[0000-0002-3266-857X]{Y. Chinone}
\affiliation{Department of Physics, University of California, Berkeley, CA 94720, USA}
\affiliation{Kavli Institute for the Physics and Mathematics of the Universe (Kavli IPMU, WPI), UTIAS, The University of Tokyo, Kashiwa, Chiba 277-8583, Japan}
\affiliation{Kavli Institute for the Physics and Mathematics of the Universe (WPI), Berkeley Satellite, the University of California, Berkeley 94720, USA}

\author{H. Miyatake}
\affiliation{Institite for Advanced Research, Nagoya University, Nagoya, Aichi, 464-8601, Japan}
\affiliation{Division of Particle and Astrophysical Science, Graduate School of Science, Nagoya University, Nagoya, Aichi, 464-8602, Japan}
\affiliation{Kavli Institute for the Physics and Mathematics of the Universe (Kavli IPMU, WPI), UTIAS, The University of Tokyo, Kashiwa, Chiba 277-8583, Japan}

\author{M. Oguri}
\affiliation{Department of Physics, The University of Tokyo, Tokyo 113-0033, Japan}
\affiliation{Kavli Institute for the Physics and Mathematics of the Universe (Kavli IPMU, WPI), UTIAS, The University of Tokyo, Kashiwa, Chiba 277-8583, Japan}
\affiliation{Research Center for the Early Universe, School of Science, The University of Tokyo, Bunkyo-ku, Tokyo 113-0033, Japan}

\author{R. Takahashi}
\affiliation{Faculty of Science and Technology, Hirosaki University, 3 Bunkyo-cho, Hirosaki, Aomori 036-8588, Japan}

\author{A. Kusaka}
\affiliation{Physics Division, Lawrence Berkeley National Laboratory, Berkeley, CA 94720, USA}
\affiliation{Department of Physics, The University of Tokyo, Tokyo 113-0033, Japan}
\affiliation{Kavli Institute for the Physics and Mathematics of the Universe (WPI), Berkeley Satellite, the University of California, Berkeley 94720, USA}
\affiliation{Research Center for the Early Universe, School of Science, The University of Tokyo, Bunkyo-ku, Tokyo 113-0033, Japan}

\author{N. Katayama}
\affiliation{Kavli Institute for the Physics and Mathematics of the Universe (Kavli IPMU, WPI), UTIAS, The University of Tokyo, Kashiwa, Chiba 277-8583, Japan}

\author[0000-0002-0400-7555]{S. Adachi}
\affiliation{Department of Physics, Kyoto University, Kyoto 606-8502, Japan}

\author[0000-0002-1571-663X]{M. Aguilar}
\affiliation{Department of Physics and Astronomy, Johns Hopkins University, Baltimore, MD 21218, USA}
\affiliation{Departamento de F\'isica, FCFM, Universidad de Chile, Blanco Encalada 2008, Santiago, Chile}

\author[0000-0001-7964-9766]{H. Aihara}
\affiliation{Department of Physics, The University of Tokyo, Tokyo 113-0033, Japan}
\affiliation{Kavli Institute for the Physics and Mathematics of the Universe (Kavli IPMU, WPI), UTIAS, The University of Tokyo, Kashiwa, Chiba 277-8583, Japan}

\author[0000-0001-7941-9602]{A. Ali}
\affiliation{Department of Physics, University of California, Berkeley, CA 94720, USA}

\author{R. Armstrong}
\affiliation{Lawrence Livermore National Laboratory, Livermore, CA 94551, USA}

\author[0000-0002-3407-5305]{K. Arnold}
\affiliation{Department of Physics, University of California, San Diego, CA 92093-0424, USA}

\author[0000-0002-8211-1630]{C. Baccigalupi}
\affiliation{International School for Advanced Studies (SISSA), Via Bonomea 265, 34136, Trieste, Italy}
\affiliation{Institute for Fundamental Physics of the Universe (IFPU), Via Beirut 2, 34151, Grignano (TS), Italy}
\affiliation{National Institute for Nuclear Physics, INFN, Sezione di Trieste Via Valerio 2, I-34127, Trieste, Italy}

\author[0000-0002-1623-5651]{D. Barron}
\affiliation{Department of Physics and Astronomy, University of New Mexico, Albuquerque, NM, 87131, USA}

\author[0000-0003-0848-2756]{D. Beck}
\affiliation{AstroParticule et Cosmologie (APC), Univ Paris Diderot, CNRS/IN2P3, CEA/Irfu, Obs de Paris, Sorbonne Paris Cit\'e, France}

\author{S. Beckman}
\affiliation{Department of Physics, University of California, Berkeley, CA 94720, USA}

\author[0000-0003-4847-3483]{F. Bianchini}
\affiliation{School of Physics, University of Melbourne, Parkville, VIC 3010, Australia}

\author{D. Boettger}
\affiliation{Instituto de Astrof\'isica and Centro de Astro-Ingenier\'ia, Facultad de F\'isica, Pontificia Universidad Cat\'olica de Chile, Av. Vicu\~na Mackenna 4860, 7820436 Macul, Santiago, Chile}

\author{J. Borrill}
\affiliation{Computational Cosmology Center, Lawrence Berkeley National Laboratory, Berkeley, CA 94720, USA}
\affiliation{Space Sciences Laboratory, University of California, Berkeley, CA 94720, USA}

\author[0000-0002-7764-378X]{K. Cheung}
\affiliation{Department of Physics, University of California, Berkeley, CA 94720, USA}

\author[0000-0002-1227-1786]{L. Corbett}
\affiliation{Department of Physics, University of California, Berkeley, CA 94720, USA}

\author[0000-0001-5068-1295]{K. T. Crowley}
\affiliation{Department of Physics, University of California, Berkeley, CA 94720, USA}

\author[0000-0001-5471-3434]{H. El Bouhargani}
\affiliation{AstroParticule et Cosmologie (APC), Univ Paris Diderot, CNRS/IN2P3, CEA/Irfu, Obs de Paris, Sorbonne Paris Cit\'e, France}

\author{T. Elleflot}
\affiliation{Department of Physics, University of California, San Diego, CA 92093-0424, USA}

\author[0000-0002-1419-0031]{J. Errard}
\affiliation{AstroParticule et Cosmologie (APC), Univ Paris Diderot, CNRS/IN2P3, CEA/Irfu, Obs de Paris, Sorbonne Paris Cit\'e, France}

\author[0000-0002-3255-4695]{G. Fabbian}
\affiliation{Department of Physics \& Astronomy, University of Sussex, Brighton BN1 9QH, UK}

\author{C. Feng}
\affiliation{Department of Physics, University of Illinois at Urbana-Champaign, 1110 W Green St, Urbana, IL, 61801, USA}

\author{N. Galitzki}
\affiliation{Department of Physics, University of California, San Diego, CA 92093-0424, USA}

\author{N. Goeckner-Wald}
\affiliation{Department of Physics, University of California, Berkeley, CA 94720, USA}

\author{J. Groh}
\affiliation{Department of Physics, University of California, Berkeley, CA 94720, USA}

\author{T. Hamada}
\affiliation{Astronomical Institute, Graduate School of Science, Tohoku University, Sendai, 980-8578, Japan}
\affiliation{High Energy Accelerator Research Organization (KEK), Tsukuba, Ibaraki 305-0801, Japan}

\author[0000-0003-1443-1082]{M. Hasegawa}
\affiliation{High Energy Accelerator Research Organization (KEK), Tsukuba, Ibaraki 305-0801, Japan}
\affiliation{SOKENDAI (The Graduate University for Advanced Studies), Shonan Village, Hayama, Kanagawa 240-0193, Japan}

\author{M. Hazumi}
\affiliation{High Energy Accelerator Research Organization (KEK), Tsukuba, Ibaraki 305-0801, Japan}
\affiliation{Kavli Institute for the Physics and Mathematics of the Universe (Kavli IPMU, WPI), UTIAS, The University of Tokyo, Kashiwa, Chiba 277-8583, Japan}
\affiliation{SOKENDAI (The Graduate University for Advanced Studies), Shonan Village, Hayama, Kanagawa 240-0193, Japan}
\affiliation{Institute of Space and Astronautical Science (ISAS), Japan Aerospace Exploration Agency (JAXA), Sagamihara, Kanagawa 252-0222, Japan}

\author{C. A. Hill}
\affiliation{Department of Physics, University of California, Berkeley, CA 94720, USA}
\affiliation{Physics Division, Lawrence Berkeley National Laboratory, Berkeley, CA 94720, USA}

\author{L. Howe}
\affiliation{Department of Physics, University of California, San Diego, CA 92093-0424, USA}

\author[0000-0001-5893-7697]{O. Jeong}
\affiliation{Department of Physics, University of California, Berkeley, CA 94720, USA}

\author{D. Kaneko}
\affiliation{Kavli Institute for the Physics and Mathematics of the Universe (Kavli IPMU, WPI), UTIAS, The University of Tokyo, Kashiwa, Chiba 277-8583, Japan}

\author[0000-0003-3118-5514]{B. Keating}
\affiliation{Department of Physics, University of California, San Diego, CA 92093-0424, USA}

\author{A. T. Lee}
\affiliation{Department of Physics, University of California, Berkeley, CA 94720, USA}
\affiliation{Physics Division, Lawrence Berkeley National Laboratory, Berkeley, CA 94720, USA}
\affiliation{Radio Astronomy Laboratory, University of California, Berkeley, CA 94720, USA}

\author{D. Leon}
\affiliation{Department of Physics, University of California, San Diego, CA 92093-0424, USA}

\author{E. Linder}
\affiliation{Space Sciences Laboratory, University of California, Berkeley, CA 94720, USA}
\affiliation{Physics Division, Lawrence Berkeley National Laboratory, Berkeley, CA 94720, USA}

\author{L. N. Lowry}
\affiliation{Department of Physics, University of California, San Diego, CA 92093-0424, USA}

\author{A. Mangu}
\affiliation{Department of Physics, University of California, Berkeley, CA 94720, USA}
\affiliation{Physics Division, Lawrence Berkeley National Laboratory, Berkeley, CA 94720, USA}

\author[0000-0003-0041-6447]{F. Matsuda}
\affiliation{Kavli Institute for the Physics and Mathematics of the Universe (Kavli IPMU, WPI), UTIAS, The University of Tokyo, Kashiwa, Chiba 277-8583, Japan}

\author[0000-0003-2176-8089]{Y. Minami}
\affiliation{High Energy Accelerator Research Organization (KEK), Tsukuba, Ibaraki 305-0801, Japan}

\author{S. Miyazaki}
\affiliation{National Astronomical Observatory of Japan, Mitaka, Tokyo 181-8588, Japan}

\author{H. Murayama}
\affiliation{Kavli Institute for the Physics and Mathematics of the Universe (Kavli IPMU, WPI), UTIAS, The University of Tokyo, Kashiwa, Chiba 277-8583, Japan}
\affiliation{Department of Physics, University of California, Berkeley, CA 94720, USA}

\author{M. Navaroli}
\affiliation{Department of Physics, University of California, San Diego, CA 92093-0424, USA}

\author[0000-0003-0738-3369]{H. Nishino}
\affiliation{High Energy Accelerator Research Organization (KEK), Tsukuba, Ibaraki 305-0801, Japan}

\author[0000-0002-6109-2397]{A. J. Nishizawa}
\affiliation{Institite for Advanced Research, Nagoya University, Nagoya, Aichi, 464-8601, Japan}

\author{A. T. P. Pham}
\affiliation{School of Physics, University of Melbourne, Parkville, VIC 3010, Australia}

\author[0000-0001-9807-3758]{D. Poletti}
\affiliation{International School for Advanced Studies (SISSA), Via Bonomea 265, 34136, Trieste, Italy}
\affiliation{Institute for Fundamental Physics of the Universe (IFPU), Via Beirut 2, 34151, Grignano (TS), Italy}
\affiliation{The National Institute for Nuclear Physics, INFN, Sezione di Trieste Via Valerio 2, I-34127, Trieste, Italy}

\author[0000-0002-0689-4290]{G. Puglisi}
\affiliation{Department of Physics, Stanford University, Stanford, CA, 94305, USA}

\author[0000-0003-2226-9169]{C. L. Reichardt}
\affiliation{School of Physics, University of Melbourne, Parkville, VIC 3010, Australia}

\author{B. D. Sherwin}
\affiliation{Department of Applied Mathematics and Theoretical Physics, University of Cambridge, Cambridge CB3 0WA, UK}
\affiliation{Kavli Institute for Cosmology Cambridge, Cambridge CB3 OHA, UK}

\author[0000-0001-7480-4341]{M. Silva-Feaver}
\affiliation{Department of Physics, University of California, San Diego, CA 92093-0424, USA}

\author[0000-0001-6830-1537]{P. Siritanasak}
\affiliation{Department of Physics, University of California, San Diego, CA 92093-0424, USA}

\author{J. S. Speagle}
\affiliation{Center for Astrophysics $\vert$ Harvard \& Smithsonian, Cambridge MA 02138, USA}

\author[0000-0002-9777-3813]{R. Stompor}
\affiliation{AstroParticule et Cosmologie (APC), Univ Paris Diderot, CNRS/IN2P3, CEA/Irfu, Obs de Paris, Sorbonne Paris Cit\'e, France}

\author[0000-0001-8101-468X]{A. Suzuki}
\affiliation{Physics Division, Lawrence Berkeley National Laboratory, Berkeley, CA 94720, USA}

\author{P. J. Tait}
\affiliation{Subaru Telescope, National Astronomical Observatory of Japan, Hilo, HI 96720, USA}

\author{O. Tajima}
\affiliation{Department of Physics, Kyoto University, Kyoto 606-8502, Japan}
\affiliation{High Energy Accelerator Research Organization (KEK), Tsukuba, Ibaraki 305-0801, Japan}

\author{M. Takada}
\affiliation{Kavli Institute for the Physics and Mathematics of the Universe (Kavli IPMU, WPI), UTIAS, The University of Tokyo, Kashiwa, Chiba 277-8583, Japan}

\author[0000-0001-9461-7519]{S. Takakura}
\affiliation{Kavli Institute for the Physics and Mathematics of the Universe (Kavli IPMU, WPI), UTIAS, The University of Tokyo, Kashiwa, Chiba 277-8583, Japan}

\author{S. Takatori}
\affiliation{SOKENDAI (The Graduate University for Advanced Studies), Shonan Village, Hayama, Kanagawa 240-0193, Japan}
\affiliation{High Energy Accelerator Research Organization (KEK), Tsukuba, Ibaraki 305-0801, Japan}

\author{D. Tanabe}
\affiliation{SOKENDAI (The Graduate University for Advanced Studies), Shonan Village, Hayama, Kanagawa 240-0193, Japan}

\author{M. Tanaka}
\affiliation{National Astronomical Observatory of Japan, Mitaka, Tokyo 181-8588, Japan}

\author{G. P. Teply}
\affiliation{Department of Physics, University of California, San Diego, CA 92093-0424, USA}

\author{C. Tsai}
\affiliation{Department of Physics, University of California, San Diego, CA 92093-0424, USA}

\author[0000-0002-3942-1609]{C. Verg\'es}
\affiliation{AstroParticule et Cosmologie (APC), Univ Paris Diderot, CNRS/IN2P3, CEA/Irfu, Obs de Paris, Sorbonne Paris Cit\'e, France}

\author[0000-0001-5109-9379]{B. Westbrook}
\affiliation{Radio Astronomy Laboratory, University of California, Berkeley, CA 94720, USA}

\author[0000-0002-5878-4237]{Y. Zhou}
\affiliation{Department of Physics, University of California, Berkeley, CA 94720, USA}

\collaboration{The \textsc{Polarbear}\ Collaboration and the Subaru HSC SSP Collaboration}

\begin{abstract}
We present the first measurement of cross-correlation between the lensing potential, reconstructed
from cosmic microwave background (CMB) {\it polarization} data, and the cosmic shear field from galaxy shapes. This measurement is made using data
from the \textsc{Polarbear}\ CMB experiment and the Subaru Hyper Suprime-Cam (HSC) survey.
By analyzing an 11~deg$^2$ overlapping region, we reject the null hypothesis at 3.5$\sigma$\ and constrain the amplitude of the { cross-spectrum} to $\widehat{A}_{\rm lens}=1.70\pm 0.48$,
where $\widehat{A}_{\rm lens}$ is the amplitude normalized with respect to the Planck~2018{} prediction, based on the flat $\Lambda$ cold dark matter cosmology.
The first measurement of this { cross-spectrum} without relying on CMB temperature measurements
is possible due to the deep \textsc{Polarbear}{} map with a noise level of ${\sim}$6\,$\mathrm{\mu K}$-arcmin,
as well as the deep HSC data with a high galaxy number density of $n_g=23\,{\rm arcmin^{-2}}$.
We present a detailed study of the systematics budget to show that residual systematics in our results are negligibly small,
which demonstrates the future potential of this cross-correlation technique.
\end{abstract}

\section{Introduction} \label{sec:intro} 

Weak lensing of the cosmic microwave background (CMB) and galaxies, referred to respectively as CMB lensing and cosmic shear, is a very powerful tool for constraining cosmology,
as it is sensitive to both the cosmic expansion and the growth of the large-scale structure (e.g., \citealt{Kilbinger:2014cea,Matilla:2017rmu}).
Furthermore, weak lensing directly probes the gravitational potential of the large-scale structure that is dominated by dark matter,
and is therefore immune to the galaxy bias uncertainty.

The constraining power of CMB lensing and cosmic shear on cosmological parameters, such as the mass fluctuation amplitude $\sigma_8$ and matter density $\Omega_m$, can be enhanced
by combining these two measurements (e.g., \citealt{P18:phi} and references therein).
In the near future, properties of dark energy (or gravity theories), dark matter, and neutrinos will be tightly constrained by such cross-correlation measurements (see e.g., \citealt{Hu:2001fb,Abazajian:2002ck,Acquaviva:2005xz,Hannestad:2006,Namikawa:2010,Abazajian:2013oma}). In addition, it has been argued that the cross-correlation between CMB lensing and cosmic shear is important to mitigate instrumental systematics inherent to these measurements~\citep{Vallinotto:2011ge,Bianchini:2014,Bianchini:2015,Liu:2016lxy,Schaan:2016ois,Abbott:2018ydy}, as cross-correlation is immune to additive instrumental biases in each lensing measurement. In the cosmic shear analysis, the calibration bias of galaxy shape measurements is one of the main sources of systematic errors, which may also be calibrated by cross-correlation. 

The cross-correlation between CMB lensing and cosmic shear has been measured by multiple experimental groups, including the Atacama Cosmology Telescope (ACT), Planck, South Pole Telescope (SPT), Sloan Digital Sky Survey (SDSS), Canada-France-Hawaii Telescope Lensing Survey (CFHTLenS),
{ CFHT Stripe 82 Survey (CS82), Red Cluster Sequence Lensing Survey (RCSLenS), Kilo Degree Survey (KiDS),} and Dark Energy Survey~(DES)~\citep{Hand:2013xua,Kirk:2015dpw,Liu:2015xfa,Singh:2016xey,Harnois-Deraps:2016huu,Harnois-Deraps:2017,Omori:2018}.
In these measurements, however, the sensitivity to CMB lensing is primarily derived from the CMB temperature data. 

One of the difficulties we are facing in CMB lensing measurements is contamination from foreground emissions in CMB lensing maps. For instance, one of the goals of future CMB instruments is to validate the shear calibration for the Large Synoptic Survey Telescope~(LSST) at the target accuracy of 0.5\%~\citep{Vallinotto:2011ge,Das:2013aia,Schaan:2016ois} by cross-correlating CMB lensing maps with the weak lensing map from LSST~\citep{S4ScienceBook,SimonsObservatory}.
Validating the LSST shear calibration requires a high CMB lensing signal-to-noise.
Ultimately, this will come from CMB polarization rather than temperature { because the $B$-mode polarization signal is mostly created from lensing while the temperature is dominated by non-lensing contributions}. Furthermore, extragalactic foregrounds cause significant biases in temperature-based lensing,
which need to be mitigated~\citep{Schaan:2019}.
One way to achieve the high lensing signal-to-noise needed and overcome the foreground issue is to resort to CMB polarization data for the lensing reconstruction~\citep{vanEngelen:2013rla,Schaan:2016ois}.

For the first time, we present the analysis of the cross-correlation between CMB lensing and cosmic shear where the CMB lensing map is reconstructed from {\it polarization} information only. This analysis is made possible by combining two deep overlapping surveys:
the CMB polarization measurement by the \textsc{Polarbear}{} experiment~\citep{Arnold_SPIE2012,Kermish_SPIE2012} and the galaxy shape measurement by Subaru Hyper Suprime-Cam~\citep[HSC;][]{Aihara:2018a}. The \textsc{Polarbear}{} CMB polarization survey is among the deepest to date, reaching 6\,$\mathrm{\mu K}$-arcmin{}.
The HSC survey is also one of the deepest wide-field optical imaging surveys, with a high galaxy number density of $n_g=23\,{\rm arcmin^{-2}}$\ for cosmic shear analyses.
The deep imaging also results in a relatively high mean redshift of these galaxies~($z_{\rm mean}=1.0$),
enhancing the overlap of the lensing kernel between CMB lensing and cosmic shear from galaxy shapes.
As such, the predicted amplitude of the cross-correlation is higher than those for the Kilo-Degree Survey~\citep{KiDS:2015} and DES~\citep{DES:2016}.
It is worth noting that our result represents the first cross-correlation measurement between HSC cosmic shear and CMB lensing~(whether in polarization or temperature).

This paper is organized as follows. In Section~\ref{sec:theory}, we briefly review the theoretical background of CMB lensing and cosmic shear.
In Section~\ref{sec:data}, we describe the data used in the analysis. In Section~\ref{sec:method},
we summarize the method to measure lensing from the CMB polarization map and the galaxy shape catalog.
We also present the results of validation tests.
We then present the cross-correlation results in Section~\ref{sec:results}, and conclude in Section~\ref{sec:conclusion}.

\section{Weak Lensing of CMB and Galaxies} \label{sec:theory} 
CMB polarization anisotropies are distorted by the gravitational potential of the large-scale structure between the CMB last scattering surface and observer~(see \citealt{Lewis:2006fu,Hanson:2009kr} for reviews). The effect of weak lensing on the CMB is well-described by a remapping of the CMB anisotropies at the last scattering surface: 
\begin{align} 
	[\widetilde{Q}\pm{\rm i}\hspace{0.05em} \widetilde{U}](\widehat{\bm{n}}) = [Q\pm {\rm i}\hspace{0.05em} U][\widehat{\bm{n}}+\bm{\nabla}\phi (\widehat{\bm{n}})] \,, 
 \end{align} 
where $\widehat{\bm{n}}$ is the pointing vector on the sky, $Q$ and $U$ ($\widetilde{Q}$ and $\widetilde{U}$) denote the primary unlensed~(lensed) Stokes parameters, and $\phi$ is the CMB lensing potential. The CMB lensing convergence, $\kappa \equiv -\bm{\nabla}^2\phi/2$, is obtained by solving the geodesic equation, yielding:
\begin{align} 
	\kappa(\widehat{\bm{n}}) = \ifstrempty{}{\int_{0}^{\chi_*} \!\! \,{\rm d}^{} \chi \, \,}{\int_{0}^{\chi_*} \!\! \frac{\,{\rm d}^{} \chi \,}{} \,} \frac{\chi_*-\chi}{\chi_*\chi} 
		\bm{\nabla}^2\Psi(\chi\widehat{\bm{n}},\chi) 
	\,, \label{Eq:cmb-kappa} 
 \end{align}
where $\chi$ is the comoving distance, $\chi_*$ denoting the comoving distance to the last scattering surface, and $\Psi(\bm{x},\chi)$ is the Weyl potential. We assume a flat universe, as we will throughout this paper. The convergence map can be reconstructed from observed CMB maps via mode coupling in CMB anisotropies induced by lensing \citep{Hu:2001kj,Okamoto:2002ik}. 

Lensing also distorts shapes of galaxy images in a galaxy survey. We can statistically measure the lensing distortion to the galaxy shapes, or the so-called shear, by { correlating} ellipticities of galaxy images~(see \citealt{Bartelmann:2001,Munshi:2008:review,Kilbinger:2014cea} for reviews). The shear field, $\gamma_1(\widehat{\bm{n}})$ and $\gamma_2(\widehat{\bm{n}})$, estimated from galaxy ellipticities, is a spin-$2$ field, and can be transformed to rotationally invariant quantities, the so-called $E$- and $B$-mode shear fields, $\gamma^E$ and $\gamma^B$, via the spin-2 transformation. Similar to the convergence field in Eq.~\eqref{Eq:cmb-kappa}, the $E$-mode shear field is related to the gravitational potential of the large-scale structure, whereas the $B$-mode shear field is generated by the vector and tensor perturbations, the post-Born correction, and other nonlinear effects (\citealt{Cooray:2002mj,Dodelson:2003bv,Cooray:2005hm,Yamauchi:2013fra}). The $B$-mode shear field is therefore expected to be very small and is usually measured as a null test.

In a survey region overlapping a CMB experiment and a galaxy survey,
a correlated signal exists between CMB lensing and cosmic shear, as they share { some of} the same large-scale structure along the line-of-sight.
Their { cross-spectrum}, $C_L^{\kappa\gamma^E}$, is of great interest in cosmological analyses,
since it is immune to additive instrumental biases inherent in these measurements.
The { cross-spectrum} from the scalar perturbations is given by (e.g., \citealt{Hu:2000ee}):
\begin{align}  
	C_L^{\kappa\gamma^E} &= \frac{2}{\pi}\ifstrempty{}{\int \!\! \,{\rm d}^{} k \, \,}{\int \!\! \frac{\,{\rm d}^{} k \,}{} \,}k^2\ifstrempty{}{\int_{0}^{\chi_*} \!\! \,{\rm d}^{} \chi \, \,}{\int_{0}^{\chi_*} \!\! \frac{\,{\rm d}^{} \chi \,}{} \,}\ifstrempty{}{\int_{0}^{\infty} \!\! \,{\rm d}^{} \chi' \, \,}{\int_{0}^{\infty} \!\! \frac{\,{\rm d}^{} \chi' \,}{} \,} 
	    \notag \\
    	&\qquad\times k^3 P_\Psi(k,\chi,\chi') S_L^\kappa(k,\chi)S_L^{\gamma^E}(k,\chi')
	\,, \label{Eq:kappaxshear} 
 \end{align}
where $L$ is the angular multipole and $k$ is the Fourier mode of the Weyl potential. The power spectrum of the Weyl potential, $P_\Psi(k,\chi,\chi')$, is defined as:
\begin{align} 
	(2\pi)^3\delta_D^3(\bm{k}-\bm{k}')P_\Psi(k,\chi,\chi') = \langle{\Psi_{\bm{k}}(\chi)\Psi^*_{\bm{k}'}(\chi')}\rangle \,, 
 \end{align}
where $\delta_D^3$ is the three-dimensional delta function, $\langle{\cdots}\rangle$ denoting the ensemble average, and $\Psi_{\bm{k}}(\chi)$ is the three-dimensional Fourier transform of the Weyl potential. The dimensionless source functions, $S_L^\kappa(k,\chi)$ and $S_L^{\gamma^E}(k,\chi)$, are given by: 
\begin{align}  
	S_L^\kappa(k,\chi) &= L(L+1)\frac{j_L(k\chi)}{k\chi}\frac{\chi_*-\chi}{\chi_*}
	\,, \label{Eq:kernel-d} \\
    S_L^{\gamma^E}(k,\chi) &= \sqrt{\frac{(L+2)!}{(L-2)!}}\frac{j_L(k\chi)}{k\chi}\ifstrempty{}{\int_{\chi}^{\infty} \!\! \,{\rm d}^{} \chi'' \, \,}{\int_{\chi}^{\infty} \!\! \frac{\,{\rm d}^{} \chi'' \,}{} \,}\frac{n(\chi'')}{\bar{n}}\frac{\chi''-\chi}{\chi''} 
	\,, \label{Eq:kernel-S} 
 \end{align} 
where $j_L(k\chi)$ is the spherical Bessel function, $n(\chi)$ is the number density distribution of galaxies as a function of the comoving distance (see below), and $\bar{n}$ is the average number density of galaxies per square arcminute. We use CAMB\footnote{Code for Anisotropies in the Microwave Background~(\url{https://camb.info/})}\citep{Lewis:1999bs} to compute the { cross-spectrum} defined above. { Here we use the fitting formula for the nonlinear matter power spectrum obtained in \citet{Takahashi:2012em}, and employ the actual redshift distribution shown later.}

\section{Data and Observations} \label{sec:data} 
\subsection{\textsc{Polarbear}} \label{sec:pb_data}
\textsc{Polarbear}{} is a CMB experiment that has been operating on the 2.5\,m Huan Tran Telescope at the James Ax Observatory
at an elevation of 5{,}190~m, in the Atacama Desert in Chile since Jan~2012.
The \textsc{Polarbear}{} receiver has an array of 1{,}274{} transition edge sensors~(TES) cooled to 0.3~K, observing the sky through lenslet-coupled double-slot dipole antennas at 150~GHz. More details on the receiver and telescope can be found in \citet{Arnold_SPIE2012} and \citet{Kermish_SPIE2012}.

\begin{figure}[!b]
\centering
\includegraphics[width=\columnwidth]{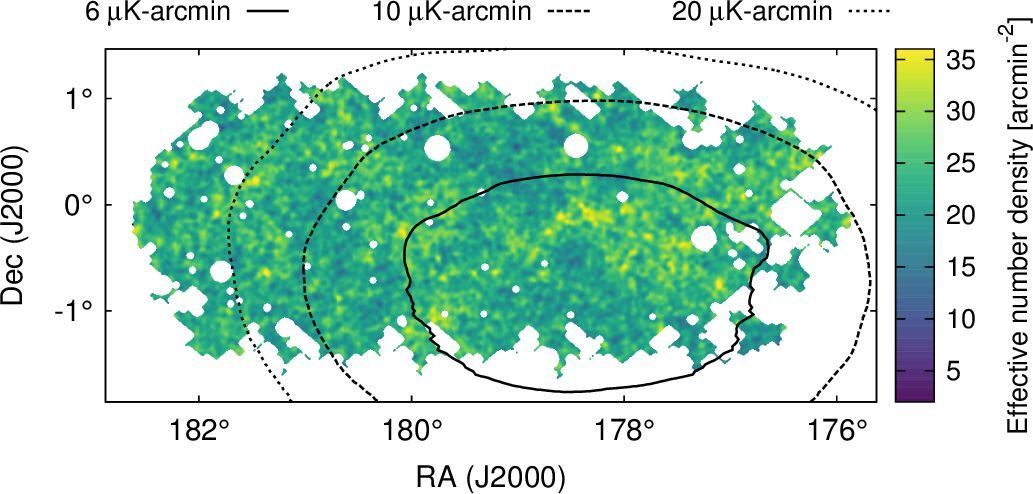}
\caption{The overlapping sky coverage of \textsc{Polarbear}{} and HSC maps in this work.
Contours show the noise level of the \textsc{Polarbear}{} CMB polarization maps.
The color map shows the effective number density of the HSC galaxy catalog.
}
\label{fig:pbxhsc_overlap}
\end{figure}

Our analysis uses data from an 11\,${\rm deg}^2$ \textsc{Polarbear}\ contiguous field that overlaps the HSC WIDE survey~(see Figure~\ref{fig:pbxhsc_overlap}). The field is centered at (RA, Dec)=({11$^\mathrm{h}$53$^\mathrm{m}$0$^\mathrm{s}$}, {$-0$$^\circ$30\arcmin}) and was observed with \textsc{Polarbear}\ for about 19 months, from 2012 to 2014.
The approximate noise level of the polarization map is 6\,$\mathrm{\mu K}$-arcmin.

The observation and map-making are described in \cite{pb2017bb}~(hereafter PB17).
Here, we use a map generated by { their} pipeline-A algorithm.
{ Based on the MASTER method~\citep{master},
the pipeline-A performs low/high-pass and azimuthal filters to remove atmospheric noise and ground pickup, respectively, prior to map-making}.
We construct an apodization window{ , $W_{\rm cmb}$,} from a smoothed inverse variance weight of the \textsc{Polarbear}{} map { as shown in Figure~\ref{fig:pbxhsc_overlap}}.
Map pixels within 3\arcmin{} of point sources are also masked. In order to reduce the $E$-$B$ leakage,
the apodization edges are modified, using the $C^2$ taper described in \citet{Grain2009}.
We multiply the $Q$/$U$ maps with this apodization window and compute the pure $B$- and $E$-modes~\citep{Smith:2005:chi-estimator}.

\subsection{HSC} \label{sec:HSC_data}
HSC is a wide-field optical imager mounted at the prime focus of the Subaru Telescope at the summit of Mauna Kea~\citep{Miyazaki:2018,Komiyama:2018}.
HSC offers a wide field-of-view~(1.77~deg$^2$), with superb image quality, and routinely $<0\farcs6$ seeing sizes,
and a fast, deep imaging capability due to the large primary mirror~(8.2~m in diameter).
As a result, HSC is one of the best instruments for weak lensing surveys. To take advantage of its survey capability,
HSC started a wide, deep galaxy imaging survey in 2014 as the Subaru Strategic Program \citep[SSP;][]{Aihara:2018a},
which includes the WIDE layer, aiming to cover 1,400~deg$^2$ of the sky down to $i_{\rm lim}\sim$26 (point source detection at 5$\sigma$) in five broad bands ($grizy$).

In this paper, we use galaxies from the first-year HSC galaxy shape catalog~\citep{Mandelbaum:2018} for the cross-correlation study.
The shape catalog includes galaxies with their $i$-band magnitudes, which are brighter than 24.5, after correcting for the Galactic extinction~\citep{Schlegel:1998}.
The shapes of these galaxies are estimated on coadded $i$-band images
with the re-Gaussianization method~\citep{Hirata2003}; this method was extensively used
in the SDSS, as its systematics are well-understood~\citep{Mandelbaum2005, Reyes2012, Mandelbaum2013}.
The shape catalog contains calibration factors for each galaxy derived from image simulations~\citep{Mandelbaum:2018b}, and generated by \textsc{GalSim}~\citep{Rowe2015}:
the shear multiplicative bias $m$~(shared among two shear components) and the additive bias for each shear component $c_1$ and $c_2$.
The following quantities are also calibrated against the image simulations: the intrinsic shape noise $e_{\rm rms}$, the estimated measurement noise ${\sigma_e}$,
and the inverse-variance weight from both $e_{\rm rms}$ and ${\sigma_e}$.
Note that we use an updated version of the shape catalog from the one originally presented in \citet{Mandelbaum:2018},
where bright stars are masked with the new ``Arcturus'' star catalog~\citep{Coupon:2018}, which is improved in comparison to the old ``Sirius'' catalog { \citep[see][for detailed discussions]{Mandelbaum:2018,Coupon:2018}}.

In this paper, we use the $13.3$~deg$^2$ HSC WIDE12H field, as it overlaps with the \textsc{Polarbear}\ survey. 
The WIDE12H field is one of six distinct fields observed from March 2014 to April 2016 for about 90 nights in total, which is a slight extension of the Public Data Release 1 \citep{Aihara:2018b}. The HSC data are reduced by the HSC pipeline~\citep{Bosch:2018}. The weighted number density of source galaxies in this field is 23.4~arcmin$^{-2}$
and its median~(mean) redshift~(see below) is $z_{\rm median} =0.88$~($z_{\rm mean}=1.0$). Figure~\ref{fig:pbxhsc_overlap} shows the overlapping sky coverage of the \textsc{Polarbear}{}
and HSC data in this paper. The overlapping sky coverage is 11.1\,${\rm deg}^2$,
where the noise level of the \textsc{Polarbear}{} polarization measurement is smaller than 20\,$\mu$K-arcmin.
\begin{figure}[!b]
\centering
\includegraphics[width=\columnwidth]{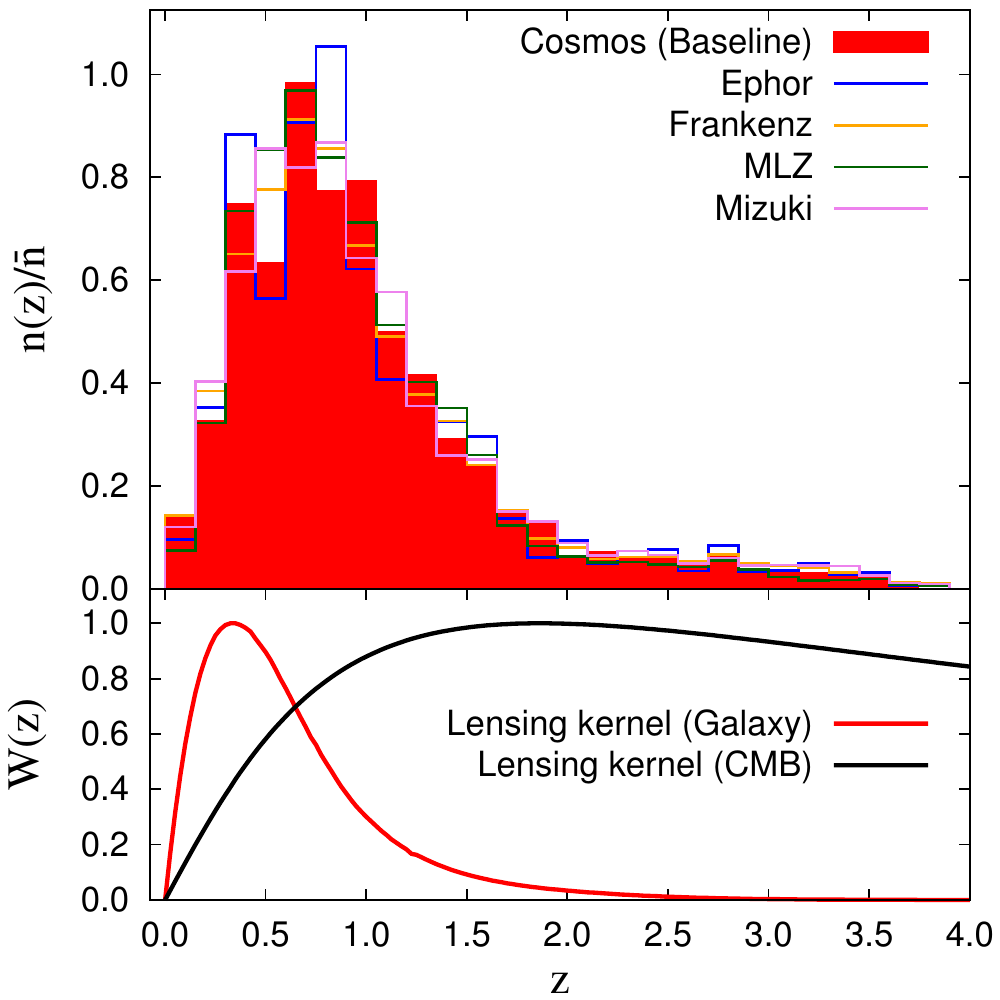}
\caption{Top:~Redshift distributions of HSC galaxies used for the cross-correlation analysis. The filled histogram shows our baseline estimate, whereas open histograms show distributions estimated from different HSC photometric redshift estimates~\citep{Tanaka:2018}.
Bottom:~Lensing kernels of the HSC galaxies~(red) and CMB~(black). The lensing kernels are normalized by their maximum values. We only show the galaxy lensing kernel for the baseline distribution.
}
\label{fig:dndz}
\end{figure}

{ 
For the baseline analysis, we use the redshift distribution of the source galaxies estimated from COSMOS 30-band photometric redshifts~\citep{Ilbert:2009},
}
which were estimated for galaxies in the COSMOS field, using 30 photometric bands spanning from ultraviolet to mid-infrared.
We reweight the redshift distribution of the COSMOS 30-band photometric redshift sample to adjust it to match our source galaxy sample on a self-organizing map
created with four colors of HSC~(More et al. in prep., see also \citealt{Miyatake:2019, Hikage:2019}).
To test the robustness of this result,
we compare the one predicated on this baseline redshift distribution,
with those obtained using several photometric redshift estimations (based solely on the four HSC colors):
``Ephor,'' ``Frankenz,'' ``MLZ,'' and ``Mizuki'' { in the WIDE12H field}~\citep{Tanaka:2018}.
For each case, the total redshift distribution of the source galaxy sample is obtained
by stacking the photometric redshift probability distribution function of each galaxy in this paper.
{ Figure~\ref{fig:dndz} shows the redshift distributions of the source galaxies derived from these methods.
The comparison of the lensing kernels between the HSC galaxies and CMB, as shown in the figure,
suggests that we typically probe the large-scale structure at $z\sim 0.5 \text{--} 1$ by our cross-correlation analysis.}

In all analyses of the HSC data, we use magnitudes corrected for the Galactic extinction. Therefore, we do not expect any cross-correlation between the dust contamination in our CMB and optical data. Although there might be a residual effect due to an imperfect correction of the Galactic extinction on galaxy magnitudes, for example, it is currently poorly understood and expected to be small compared to the noise level of our cross-correlation signal.

\subsection{Simulated Data} \label{sec:simulation}
We create simulated data to estimate the covariance and to perform validation tests.
The mock simulations are based on the all-sky ray-tracing simulations generated by \citet{Takahashi:2017}, and in each one,
they generate both CMB and galaxy lensing signals.
We then add realistic noise, following noise properties of each survey as described below.
From an all-sky ray-tracing simulation, we randomly cut out areas corresponding to the HSC WIDE12H geometry to create many { independent} realizations.
In total, we generate 100~WIDE12H field realizations from the single all-sky realization.
{ \citet{Takahashi:2017} confirmed that such non-overlapping regions taken from a single full-sky map are mutually independent.}

We add HSC source galaxies to the ray-tracing simulation following the prescription described in \citet{Oguri:2018}. We start with the real HSC galaxy catalog in order to simulate survey features such as the survey geometry, the inhomogeneity of the galaxy distribution,
and galaxy properties including redshifts{ , which is randomly drawn from the photometric redshift probability density function of each galaxy estimated by MLZ,} and intrinsic shapes.
The galaxy positions and redshifts are maintained unchanged but their shapes are randomly rotated to remove the weak lensing shear associated with the real data.
By doing so, we can also preserve the shot-noise originating from galaxy intrinsic shapes and pixel noises.
We then add the simulated cosmic shear field derived from the ray-tracing simulation to each rotated galaxy shape to create a mock catalog. 

For CMB, we generate Monte Carlo~(MC) simulations, having similar properties to the \textsc{Polarbear}\ data, by scanning a lensed CMB $Q$/$U$ polarization map
from the all-sky simulation described above. 
We then add a random noise to the simulated detector timestream, where the variance of the noise is equivalent to that measured from the data.

\section{Lensing Reconstruction and Cross-correlation Methods}
\label{sec:method} 
In this section, we describe our method for the cross-correlation analysis.
The lensing reconstruction and { cross-spectrum} estimator are described in Section~\ref{sec:estimator}.
We also describe the validation tests for \textsc{Polarbear}{} CMB lensing in Section~\ref{sec:validation:cmb} and HSC cosmic shear in Section~\ref{sec:validation:HSC}.

Since the auto spectra of CMB lensing and cosmic shear are validated in PB17 and \citet{Hikage:2019}, \citet{Oguri:2018}, and \citet{Mandelbaum:2018}, respectively,
we focus here on the validation tests for the { cross-spectrum} between CMB lensing and cosmic shear.

\subsection{Estimators} \label{sec:estimator}

\subsubsection{CMB Lensing Convergence} \label{sec:estimator:cmb}

Reconstruction methods of the CMB lensing convergence have been developed by multiple CMB collaborations, including ACT \citep{ACT16:phi}, BICEP/Keck Array \citep{BKVIII}, Planck \citep{P18:phi}, \textsc{Polarbear}~\citep{PB14:phi,PB14:phixCIB}, and SPT \citep{SPT:phi,Story:2014hni}. 

In this paper, we first apply the diagonal inverse-variance filter defined in Eq.~(17) of \citet{BKVIII} to the $E$- and $B$-modes obtained in Section~\ref{sec:pb_data}. The unnormalized quadratic estimator for the lensing convergence is obtained by convolving two CMB $E$-modes ($EE$ estimator) or $E$- and $B$-modes ($EB$ estimator) \citep{Hu:2001kj}:
\begin{align} 
	\overline{\kappa}^{XY}_{\bm{L}} &= \ifstrempty{(2\pi)^2}{\int \!\! \,{\rm d}^{2} \bm{\ell} \, \,}{\int \!\! \frac{\,{\rm d}^{2} \bm{\ell} \,}{(2\pi)^2} \,} w^{XY}_{\bm{\ell},\bm{L}} \overline{X}_{\bm{\ell}}\overline{Y}_{\bm{L}-\bm{\ell}} \,,
 \end{align}
where $\overline{X}$ and $\overline{Y}$ are either $E$- or $B$-modes filtered by the diagonal inverse-variance. The weight functions are given by \cite{Hu:2001kj}:
\begin{align} 
	w^{EE}_{\bm{\ell},\bm{L}} &= [C^{EE}_\ell \bm{L}\cdot\bm{\ell} + C^{EE}_{|\bm{L}-\bm{\ell}|}(\bm{L}-\bm{\ell})\cdot\bm{L}]\cos 2(\varphi_{\bm{\ell}}-\varphi_{\bm{L}-\bm{\ell}}) \,, \\
	w^{EB}_{\bm{\ell},\bm{L}} &= C^{EE}_\ell \bm{L}\cdot\bm{\ell} \sin 2(\varphi_{\bm{\ell}}-\varphi_{\bm{L}-\bm{\ell}}) \,, 
 \end{align}
where $C^{EE}_{\ell}$ is the lensed CMB $E$-mode spectrum \citep{Hanson:2010rp}, and $\varphi_{\bm{\ell}}$ is { the angle between $\bm{\ell}$ and the $x$-axis}. We use the CMB multipole range of $500\leq \ell\leq 2700$ in our baseline analysis. The larger multipoles are removed to avoid beam uncertainties and systematic biases due to astrophysical foregrounds such as radio sources \citep{vanEngelen:2013rla}. { The multipoles at $\ell<500$ is not used because they are not validated in PB17}. We then obtain our best estimate of the CMB lensing convergence as:
\begin{align} 
	\widehat{\kappa}^{XY}_{\bm{L}} = A^{XY}_L(\overline{\kappa}^{XY}_{\bm{L}}-\langle{\overline{\kappa}^{XY}_{\bm{L}}}\rangle) \,.
 \end{align}
The mean field, $\langle{\overline{\kappa}^{XY}_{\bm{L}}}\rangle$, is sourced from, for example, masking, inhomogeneous map noise, point sources, and the asymmetric beam \citep{Hanson:2010rp,Namikawa:2012}, and is non-zero, even if we use polarization-only estimators (e.g., \citealt{P13:phi}).
We estimate the mean-field bias from simulation, and the bias is found to be much smaller than the lensing signal in our case.
The normalization, $A^{XY}_L$, is computed by following \citet{PB14:phixCIB}.
Finally, the minimum variance estimator~(MV) is obtained by combining the $EE$ and $EB$ estimators \citep{Hu:2001kj}. 

\subsubsection{Cosmic Shear}

The shear field at a pixel $\widehat{\bm{n}}$ is estimated from the galaxy shape catalog as \citep{Mandelbaum:2018}:
\begin{align} 
	\gamma_j(\widehat{\bm{n}}) &= \frac{1}{\sum_{i\in g_{\widehat{\bm{n}}}}w_i}\frac{1}{1+m}\sum_{i\in g_{\widehat{\bm{n}}}}w_i\left[\frac{e_{j,i}}{2R} - c_{j,i}\right] \,,
    \label{Eq:shear}
 \end{align}
where $e_{j,i}$ ($j=1,2$) is the ellipticity of the $i$-th galaxy, $c_{j,i}$ is the additive bias, $w_i$ is the inverse-variance weight, and $\sum_{i\in g_{\widehat{\bm{n}}}}$ is the summation over all galaxies, falling within pixel $\widehat{\bm{n}}$. { The averaged multiplicative bias, $m$, and the shear responsivity, $R$, are derived as:} 
\begin{align} 
	m &= \frac{\sum_{i\in g_{\rm all}} w_i m_i}{\sum_{i\in g_{\rm all}}w_i}  \,, \\
    R &= 1 - \frac{\sum_{i\in g_{\rm all}} w_i e^2_{{\rm rms},i}}{\sum_{i\in g_{\rm all}}w_i} \,. 
 \end{align}
Here, { $m_i$ is the multiplicative bias,} $\sum_{i\in g_{\rm all}}$ is the summation over all galaxies for the cross-correlation analysis, and $e_{{\rm rms},i}$ is the root-mean square of intrinsic ellipticities. { The shear maps, $\gamma_1$ and $\gamma_2$, are} then multiplied by a window function constructed from the weight, $W_{\rm gal}(\widehat{\bm{n}})=\sum_{i\in g_{\widehat{\bm{n}}}}w_i$, and transformed to $E$- and $B$-mode shear { fields} as:
\begin{align} 
	\gamma^E_{\bm{L}}\pm{\rm i}\hspace{0.05em}\gamma^B_{\bm{L}} = \ifstrempty{}{\int \!\! \,{\rm d}^{2} \widehat{\bm{n}} \, \,}{\int \!\! \frac{\,{\rm d}^{2} \widehat{\bm{n}} \,}{} \,} \hspace{0.1em}{\rm e}^{-{\rm i}\hspace{0.05em}\bm{L}\cdot\widehat{\bm{n}}} \hspace{0.1em}{\rm e}^{\pm 2{\rm i}\hspace{0.05em}\varphi_{\bm{L}}}W_{\rm gal}(\widehat{\bm{n}})[\gamma_1\pm{\rm i}\hspace{0.05em}\gamma_2](\widehat{\bm{n}}) \,.
 \end{align}

\subsubsection{Cross-spectrum}

The binned { cross-spectrum} is obtained by cross-correlating the { CMB} lensing convergence and the $E$-mode shear field derived above. Since the { cross-spectrum} is a correlation between two CMB and one cosmic shear maps, the { cross-spectrum} is further divided by $\int {\rm d}^2\widehat{\bm{n}}\, W^2_{\rm cmb}(\widehat{\bm{n}})W_{\rm gal}(\widehat{\bm{n}})$ to correct the normalization due to the apodization window, { where $W_{\rm cmb}$ is the CMB apodization window defined in Section~\ref{sec:pb_data}}.
The number of multipole bins is $9$ and the multipole range of the output power spectrum is $100\leq L\leq 1900$. The lower limit of $L$ is set by the size of the survey region, and the higher limit of $L$ is set
because signal-to-noise ratios above $L=1900$ are negligibly small.
{ Before unblinding, we confirmed that the measured spectrum from the realistic MC simulation reproduces the input power spectrum within the simulation error.} 

\subsection{Validation Tests: CMB} \label{sec:validation:cmb}
We describe a suite of data-split null tests and instrumental systematics to validate CMB datasets in the { cross-spectrum}.
\subsubsection{Data-split Null Tests}
\label{sec:data-split_null-tests}
In order to validate the \textsc{Polarbear}{} data and analysis in the cross-correlation with the HSC data, we perform a suite of null tests. These validation tests are essentially the extension of those described in PB17 to the cross-correlation, in which we iteratively run the null-test framework until a set of predefined criteria is passed. 

For each null test, we reconstruct two lensing maps, $\kappa^A$ and $\kappa^B$, one from each data split. The reconstructed lensing maps are then cross-correlated with the HSC shear map to obtain a null spectrum for the difference between the two cross-spectra, $C_L^{\kappa^A\gamma^E}-C_L^{\kappa^B\gamma^E}$. To evaluate the statistical significance, we repeat the same calculation using the simulated CMB maps, but with the actual HSC shear data. 

The null tests are performed for several splits of interest for the \textsc{Polarbear}{} data,
which are identified to be sensitive to various sources of systematic contaminations or miscalibrations.
We perform 12 null tests in total, and the correlations among these null tests are noted in the analysis
by running the same suite of null tests on noise-only MC simulations.
{ Four tests divide the data by the observation period such as the first season data and the second season data.
Three tests target effects that depend on the telescope pointing/scan such as data taken at high or low elevation.
Three tests divide based on the proximity of the main or sidelobe beams to the Sun and Moon.
Two tests divide the data by focal-plane pixels based on susceptibility to instrumental effects.}
Details of the 12 null tests are described in \citet[][hereafter PB14]{PB14:BB} and PB17.

\begin{figure}
\centering
\includegraphics[width=0.9\columnwidth]{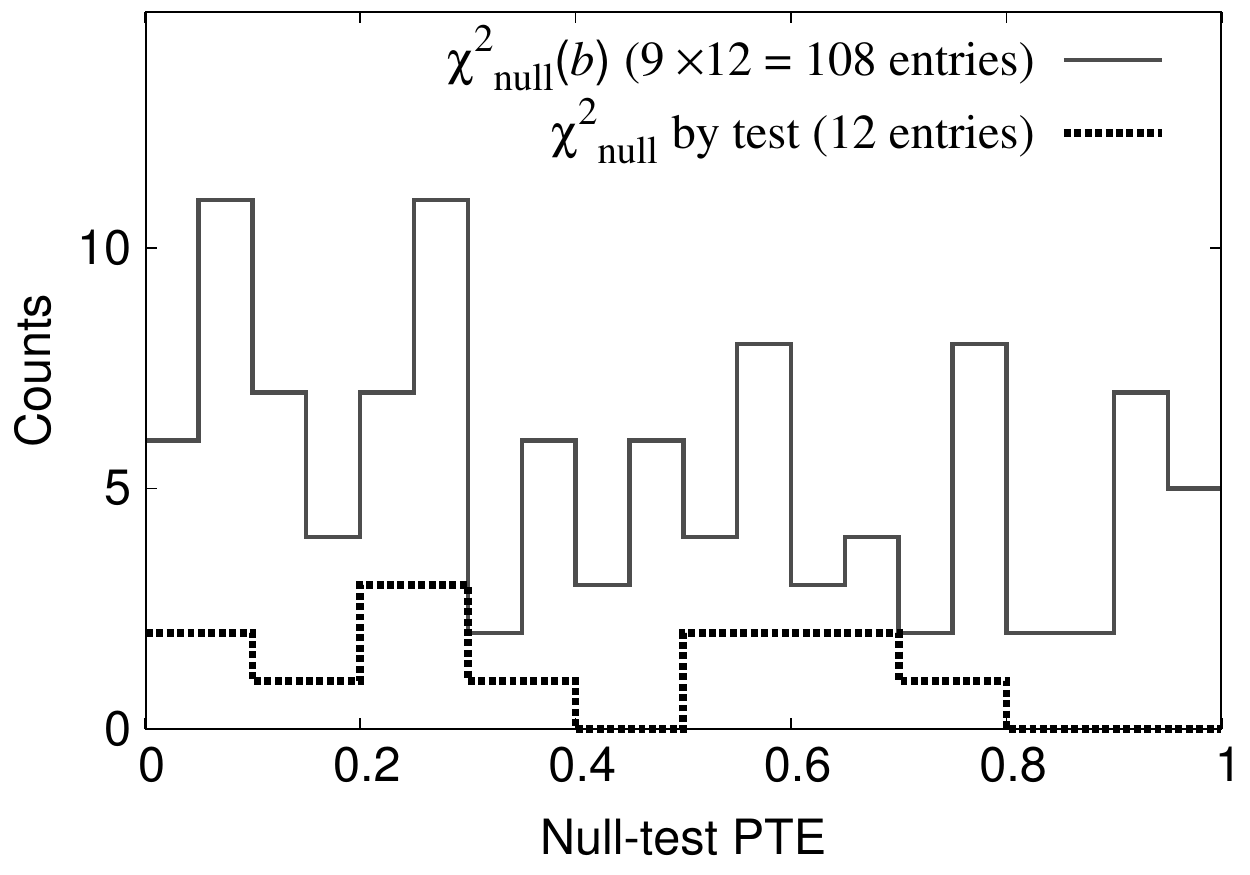}
\caption{Null-test PTE distributions of $\chi^2_{\rm null}(b)$ and $\chi^2_{\rm null}$ by test~(dotted line). Both distributions are consistent with the expectation from the uniform distribution~(see also Table~\ref{tab:pte_summary_data-split}).}
\label{fig:pbxhsc_data-split_pte_dist}
\end{figure}

We also adopt the same null-test statistics as defined in PB17. For each band power bin~$b$, we calculate the statistic $\chi_{\rm null}(b)\equiv \hat{C}_b^{\rm null}/\sigma_b$, where $\sigma_b$ is a MC-based estimate of the standard deviation of the null spectra,
and its square $\chi_{\rm null}^2(b)$. The $\chi_{\rm null}(b)$ is sensitive to a systematic bias in the null spectra,
whereas the $\chi_{\rm null}^2(b)$ is more sensitive to outliers and excess in the variance.
In order to investigate possible systematic contaminations or miscalibrations affecting a specific null-test data split,
we calculate the sum of the $\chi_{\rm null}^2(b)$ over $100 \le L \le 1900$~(``$\chi^2_{\rm null}$ by test'').
{ We require each set of probability to exceeds~(PTEs) from the $\chi_{\rm null}^2(b)$ and the $\chi_{\rm null}^2$ by test to be consistent with a uniform distribution.
We have evaluated it by using a Kolmogorov-Smirnov~(KS) test, to be equal to or greater than 0.05}.
We find these distributions consistent with the uniform distribution.
Figure~\ref{fig:pbxhsc_data-split_pte_dist} and Table~\ref{tab:pte_summary_data-split} show PTE distributions of $\chi^2_{\rm null}(b)$ and $\chi^2_{\rm null}$ by test,
and the PTEs of the KS test.
\begin{table}
\begin{center}
\caption{\label{tab:pte_summary_data-split}PTEs from the data-split null tests.}
\begin{tabular}{rll}\hline
& Type & PTE \\\hline
   & KS test of ${\rm PTE}_{\chi^2_{\rm null}(b)}$ & 0.07 \\
   & KS test of ${\rm PTE}_{\chi^2_{\rm null}\text{ by test}}$ & 0.63 \\\hline
(1) & Average of $\chi_{\rm null}(b)$ & 0.88 \\
(2) & Extreme of $\chi^2_{\rm null}(b)$ & 0.44 \\
(3) & Extreme of $\chi^2_{\rm null}$ by test & 0.16 \\
(4) & Total $\chi^2_{\rm  null}$ & 0.10 \\\hline
\end{tabular}
\end{center}
\end{table}

In order to search for different manifestations of systematic contaminations, we also create the same test statistics based on these quantities described in PB17.
The four test statistics are PTEs from (1)~the average value of $\chi_{\rm null}$,
(2)~the extreme value of the $\chi^2_{\rm null}$ by bin,
(3)~that by test, and (4)~the total $\chi^2_{\rm null}$ summed by the 12 null tests. 
In each case, the result from the data is compared to the result from simulations to calculate PTEs, as Table~\ref{tab:pte_summary_data-split} summarizes the PTEs. 

Finally, by comparing the most significant outlier from the four test statistics to those of the MC simulations, we obtain a PTE of 0.24. In all tests,
we find no evidence for systematic contaminations or miscalibrations in the \textsc{Polarbear}{} dataset correlated with the HSC dataset.

\subsubsection{Instrumental Systematics}
\label{sec:inst_systematics}
\begin{figure*}
\centering
\includegraphics[width=120mm,clip]{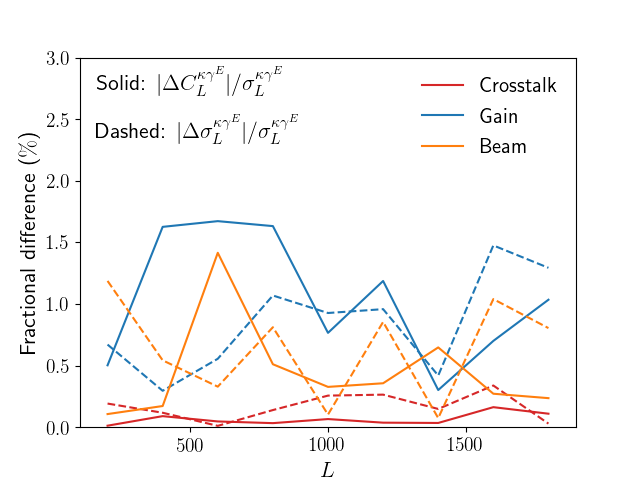}
\caption{Upper limits of the impact of the \textsc{Polarbear}{} CMB systematics on the CMB-galaxy lensing { cross-spectrum}~(solid), $|\Delta C_L^{\kappa\gamma^E}|/\sigma_L^{\kappa\gamma^E}$, and its standard deviation (dashed), $|\Delta \sigma^{\kappa\gamma^E}_L|/\sigma_L^{\kappa\gamma^E}$. We consider systematic effects: crosstalk in the multiplexed readout~({ ``crosstalk''}), the total effect of the drift of gains between two consecutive thermal source calibrator measurements and the relative gain-calibration uncertainty between the two detectors in a focal-plane pixel~(``gain''), total effect of the differential beam ellipticity, differential beam size, and differential pointing between the two detectors in a focal-plane pixel~(``beam'').  The systematic effects are combined in quadrature to derive the fractional difference of the systematics-free spectrum,
as positively defined. We find there is no preference in the sign of the fractional difference,
indicating that estimates are dominated by statistical fluctuation of MC realizations and are the conservative upper limits.
In terms of $A_{\rm lens}$, the upper limit corresponds to $1.3\%${} of the fiducial amplitude.}
\label{fig:pb-sys-spec}
\end{figure*}
We study the impact of uncertainties in the instrument model of \textsc{Polarbear}{} on the lensing auto and cross-spectra by producing a simulated signal-only data set in a time domain where the signal is modeled with lensed CMB simulations, obtained by \mbox{LensPix}\footnote{\url{https://cosmologist.info/lenspix/}}
with instrumental effects added on the fly.
With this simulation setup, systematic errors in auto spectra contains both multiplicative and additive bias in CMB lensing measurement, allowing us to put a conservative upper limit on the multiplicative component.  On the other hand, this simulation has zero expectation value in the cross power, while containing fiducial power in each of the CMB and weak lensing maps; this is an appropriate setup for estimating the additive component, whose estimate can depend on the signal power of each map.
Here, we investigated six instrumental systematics effects:
crosstalk in the multiplexed readout,
drift of the gains between two consecutive thermal source calibrator measurements,
differential beam ellipticity,
differential beam size,
relative gain-calibration uncertainty between the two detectors in a focal-plane pixel,
and differential pointing between the two detectors in a focal-plane pixel.
Details of these systematic effects and systematics simulations are described in PB14 and PB17.
{ These} contamination{ s} are found not to bias lensing auto spectrum significantly, putting an upper limit on multiplicative bias.
The limit corresponds to $0.6\%${} of fiducial lensing amplitude in the $A_{\rm lens}$ measurement~(\textsc{Polarbear}{} collaboration et al., in prep.).

In order to explicitly check the impact of the instrumental systematics on { cross-spectrum},
we reconstructed the CMB maps with \textsc{Polarbear}\ pipeline-A and the corresponding CMB lensing convergence maps from the simulated data set.
These maps are cross-correlated with the HSC mock data as described in Section~\ref{sec:simulation}.
Figure~\ref{fig:pb-sys-spec} shows { the} impact of the CMB instrumental systematics on the cross-spectra.
As expected, all the systematics and their variances are negligibly small, compared to the statistical errors estimated from the MC simulations.
Specifically, we find upper limits of ${\sim} 1$\% level on the instrumental systematic errors compared to the statistical errors { for most cases}.
We therefore find no evidence for significant contaminations from the CMB instrumental systematics in the cross-correlation analysis.
The upper limit corresponds to $1.3\%$, in terms of $A_{\rm lens}$, when compared to fiducial amplitude.

While our estimate of the systematics is for the \textsc{Polarbear}{} instrument, future CMB instruments aim to achieve similar,
if not better, levels of systematics.
We detect no significant systematic error and the upper limit presented here is dominated by the statistical uncertainty of MC simulations.
The upper limit is already comparable to the goals of Simons Observatory and CMB-S4,
which calibrate the shear bias of LSST to ${\sim} 0.5\%$  accuracy~\citep{Schaan:2016ois,S4ScienceBook,SimonsObservatory}.

\subsection{Validation Tests: Shear} \label{sec:validation:HSC}
\begin{figure*}
\begin{center}
\includegraphics[width=120mm,clip]{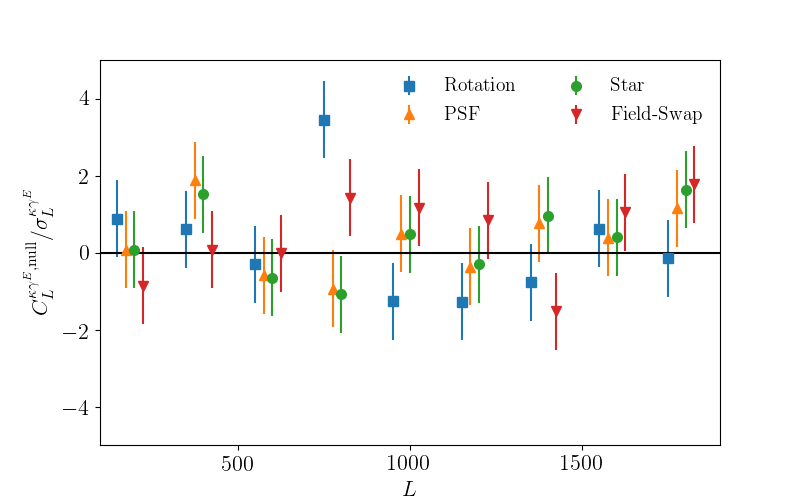}
\caption{
cross-spectra between the HSC null test maps and the real \textsc{Polarbear}{} lensing map.
We consider HSC null test maps derived by randomly rotating ellipticities of real HSC galaxies (``Rotation''), from star ellipticities (``Star''), from PSF ellipticities (``PSF''),
which are measured in another HSC field (``Field Swap''). The cross-spectra are normalized by their statistical uncertainties.}
\label{fig:hsc-null-spec}
\end{center}
\end{figure*}
We perform four validation tests for the shear map derived from the HSC data,
by cross-correlating the CMB lensing map with the following null test maps, created in the same manner as the real shear map:
\begin{itemize}
\item \textbf{Rotation}: a map from randomly rotated ellipticities of galaxies in the HSC WIDE12H field to remove the cosmic shear signal,
\item \textbf{Star}: a map from ellipticities of stars for reconstructing the Point Spread Function~(PSF), again in the HSC WIDE12H field,
\item \textbf{PSF}: a map from PSFs reconstructed at the star position, and
\item \textbf{Field Swap}: a shear map measured in another field, not overlapping with the WIDE12H field.
\end{itemize}
We expect null signals for all four cases, since these maps do not have any physical correlation with the CMB lensing map.

For the Rotation test, we measure cross-spectra without correcting for multiplicative and additive biases, i.e., we set $m_i=0$ and $c_i=0$ in Eq.~\eqref{Eq:shear}.
This ignorance of the multiplicative bias does not affect our validation test. For the Star and PSF tests, we measure their cross-power spectra with $w_i=1$, $m_i=0$, $e_{{\rm rms},i}=0$, and $c_i=0$. The equal weight is derived from the fact that all stars in this test have similar signal-to-noise ratio, which is also the case for PSFs. The zero RMS ellipticity is derived from the fact that stars and PSFs have approximately zero ellipticity, on average.

We estimate the covariance and PTEs as follows. We first generate simulations by randomly rotating ellipticities of galaxies, stars, and PSFs in the real WIDE12H field data for the Rotation, the Star, and the PSF tests, respectively. We then measure cross-spectra in a consistent way with the measurement described above. Based on 100 realizations of the HSC maps with randomly rotated ellipticities, we estimate the covariance and use it to compute PTEs. 

For the Field Swap test, we use another patch of the HSC first-year shear catalog, GAMA09H, which is located in RA of ${\sim}9$h. Since there is no overlap of the footprints between the \textsc{Polarbear}{} field and the GAMA09H field, 
there is no cross-correlation between these data.
We compute the shear map of the GAMA09H field using the same method as described in Section~\ref{sec:estimator} with the calibration of multiplicative and additive bias. In order to estimate the covariance and PTEs, we use the 100 realizations of simulations similar to those described in Section~\ref{sec:simulation} but remove to match the GAMA09H area.
Note that the mock shear catalogs contain the same calibration bias as in the real HSC shear catalog. 

Figure~\ref{fig:hsc-null-spec} shows the cross-spectra between the HSC null test maps and the real CMB lensing map. The results of these null tests are also summarized in Table~\ref{table:hscnull}. We find no evidence for systematic errors from this analysis.

\begin{table}[htbp]
\begin{center}
\caption{
Results of the HSC shear null tests. 
}
\label{table:hscnull} \vspace{0.5\baselineskip}
\begin{tabular}{lcc} \hline
 & $\chi$-PTE & $\chi^2$-PTE \\ \hline 
Rotation & 0.52 & 0.10 \\ 
Star & 0.26 & 0.43 \\ 
PSF & 0.46 & 0.49 \\ 
Field Swap & 0.20 & 0.33 \\ \hline
\end{tabular}
\end{center}
\end{table}

\subsection{Blind Analysis}
We adopt a blind analysis policy, in which the { cross-spectrum} is revealed only after the data pass a series of null tests
and systematic error checks as described in Section~\ref{sec:validation:cmb} and \ref{sec:validation:HSC}.
For the HSC data, we prepare three shape catalogs with different multiplicative biases, each of which has a different, blinded offset.
For details of the blinding strategy, see \citet{Hikage:2019}.
For the \textsc{Polarbear}{} data, the null tests and the possible sources of instrumental systematic errors are finalized
before the { cross-spectrum} is examined, in order to motivate a comprehensive validation of the dataset and to avoid an observer bias in the analysis.

\section{Results} \label{sec:results} 
\begin{figure*}
\centering
\includegraphics[width=120mm,clip]{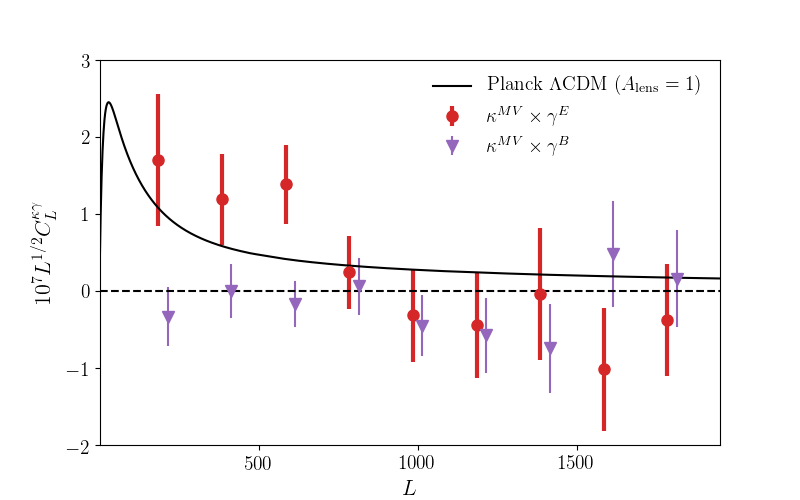}
\caption{
The cross-spectra between the CMB lensing convergence from \textsc{Polarbear}\ and the cosmic shear from HSC.
The CMB lensing map is obtained from the optimal combination of the $EE$ and $EB$ estimators~(MV).
We show the { cross-spectrum} between the HSC shear $B$-mode and the CMB lensing convergence, consistent with zero as expected.
The black solid line shows the theoretical prediction, assuming the Planck~2018{} best-fit cosmological parameters for the flat $\Lambda$CDM model. }
\label{fig:main}
\end{figure*}

\begin{figure}
\centering
\includegraphics[width=9cm,clip]{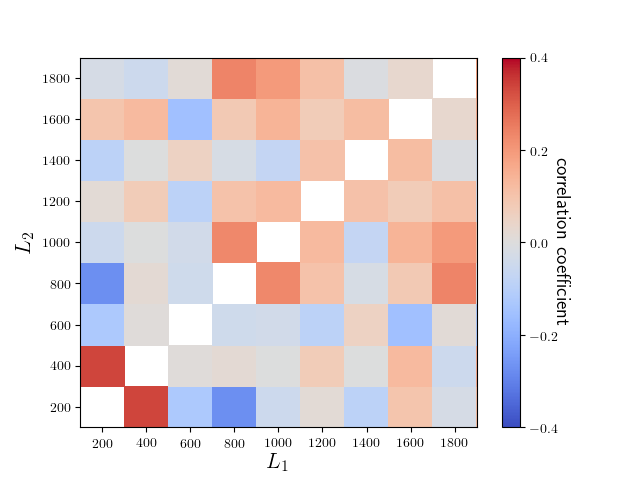}
\caption{
Correlation coefficients of the { cross-spectrum} between CMB lensing and cosmic shear, estimated from $100$ realizations of simulations.}
\label{fig:correlation}
\end{figure}

Figure~\ref{fig:main} shows the angular { cross-spectrum} between \textsc{Polarbear}{} CMB lensing and HSC cosmic shear.
We show the { cross-spectrum} measured spectra using the optimal combination of the $EE$ and $EB$ estimators~(MV). Since our CMB $B$-mode map is very deep, the power spectrum from the $EB$ estimator is less noisy than that from the $EE$ estimator. We also find that the { cross-spectrum} between the HSC $B$-mode shear and the CMB lensing convergence is consistent with zero~($\chi\text{-PTE}$ and $\chi^2\text{-PTE}$ of 0.26 and 0.68, respectively) as expected. { Figure~\ref{fig:correlation} shows the correlation coefficients among different multipole bins of the { cross-spectrum}, defined as;
\begin{align} 
	R_{bb'} = \frac{{\bm {\mathrm{Cov}}} _{bb'}}{\sqrt{{\bm {\mathrm{Cov}}} _{bb}}\sqrt{{\bm {\mathrm{Cov}}} _{b'b'}}} \,. 
 \end{align}
Here, ${\bm {\mathrm{Cov}}} _{bb'}=\langle{C_bC_{b'}}\rangle-\langle{C_b}\rangle\langle{C_{b'}}\rangle$ is the covariance of the binned cross-power spectrum. The correlation coefficient between the first and second bandpowers is ${\sim}0.4$, and that between the first and fourth bandpowers is ${\sim}-0.3$. Most of the correlation coefficients is consistent with zero within statistical uncertainty~(${\sim}10\%$) from the finite number of the MC realizations.
}

To see the consistency of our { cross-spectrum} measurement with the Planck $\Lambda$-dominated cold dark matter~($\Lambda$CDM) cosmology,
we estimate the amplitude of the { cross-spectrum} by a weighted mean over multipole bins~(\citealt{BKVIII}):
\begin{align} 
	\widehat{A}_{\rm lens} = \frac{\sum_b a_b A_b}{\sum_b a_b}  \,. \label{Eq:AL}
 \end{align}
The $A_b$ is the relative amplitude of the power spectrum compared with a fiducial power spectrum for the Planck $\Lambda$CDM cosmology, $C_b^{\rm f}$, i.e., $A_b \equiv C_b/C_b^{\rm f}$. The weights, $a_b$, are taken from the bandpower covariance as:
\begin{align} 
	a_b = \sum_{b'} C^{\rm f}_b {\bm {\mathrm{Cov}}} _{bb'}^{-1} C^{\rm f}_{b'}  \,.
 \end{align}
The fiducial bandpower values and their covariances, including off-diagonal correlations between different multipole bins, are evaluated from the simulations (see Section~\ref{sec:simulation}).
In our baseline analysis, we assume the $\Lambda$CDM cosmology with the Planck~2018\ best-fit parameters~(TT,TE,EE+lowE+lensing).


The amplitude estimated from the observed { cross-spectrum} is $\widehat{A}_{\rm lens}=1.70\pm 0.48$,\footnote{Our simulations assume the \mbox{WMAP-9} best-fit cosmology, whereas the baseline analysis of the amplitude is measured against the Planck~2018\ best-fit cosmology~(``TT,TE,EE+lowE+lensing'' in \citealt{P18:main}).
This leads to a small change in the mean and scatter of the amplitude parameter.
We correct this discrepancy by scaling the simulated { cross-spectrum} at each realization as $C^i_b\times (C^{\rm f}_b/\langle{C^i_b}\rangle)$.
The variance of the amplitude of simulations is scaled by a value estimated from analytic calculations of cross-spectra in Planck and \mbox{WMAP-9} cosmologies, using $n_g=23\,{\rm arcmin^{-2}}$, $e_{\rm rms}=0.4$, a 6\,$\mathrm{\mu K}$-arcmin\ CMB white noise, and a $3\farcm5$ Gaussian beam.}
corresponding to the detection of a non-zero cross-correlation at 3.5$\sigma${} significance.
Here, the quoted error is the standard deviation of $A_{\rm lens}$ obtained from the MC simulations.
{ The MC error in the covariance changes the cross-spectrum amplitude by only $\Delta \widehat{A}_{\rm lens}=\pm 0.06$.}
The high detection significance is in part because of the central value fluctuated high; for a fiducial value of $A_{\rm lens}=1$,
the expected signal-to-noise ratio is $S/N\sim 2$.
The PTE of the spectrum with respect to the fiducial Planck $\Lambda$CDM cosmology is $66$\%.

\begin{figure*}[!t]
 \centering
 \includegraphics[width=0.75\textwidth]{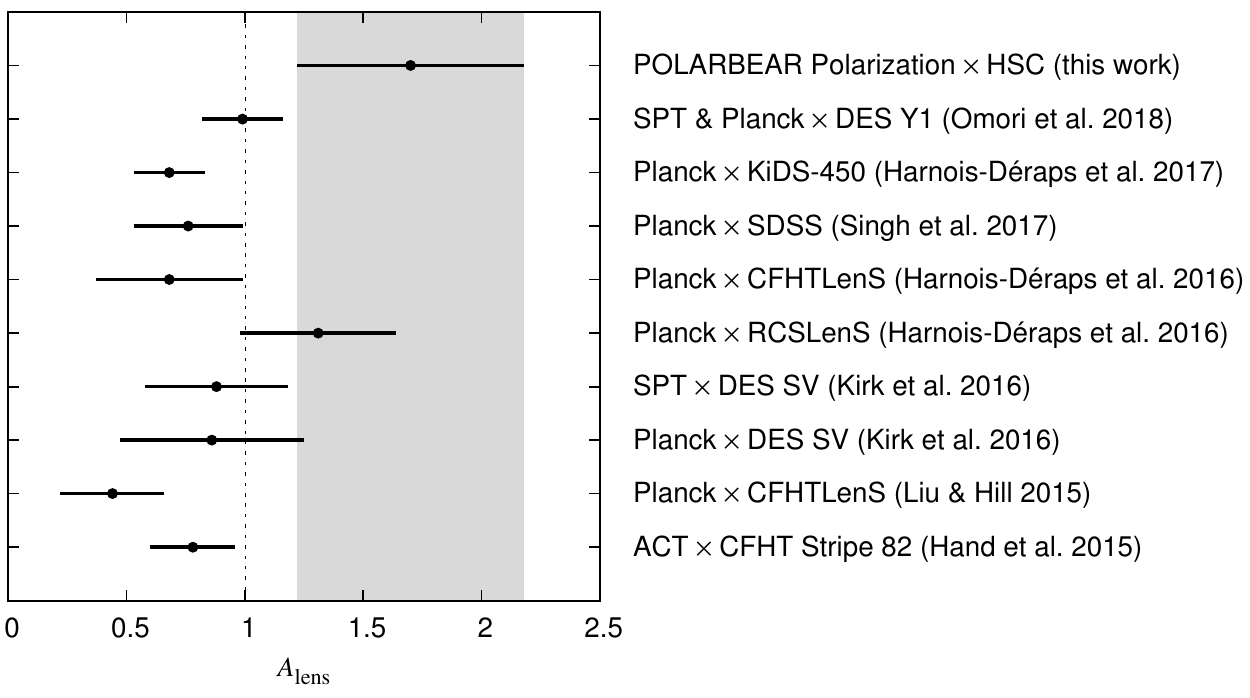}
 \caption{\label{fig:Alens_comp}
The 1$\sigma$ confidence interval on $A_{\rm lens}$ from the cross-correlation analysis between \textsc{Polarbear}{} and HSC data in the Planck $\Lambda$CDM model, as well as those from the literature.
The redshift distributions of source galaxies are different among these measurements, 
spanning from $z_{\rm mean}\sim 0.35$ \citep{Singh:2016xey} to $z_{\rm mean}\sim 1.0$ (this work).
Further details of the redshift distributions can be found in the literature~\citep{Omori:2018,Harnois-Deraps:2017,Singh:2016xey,Harnois-Deraps:2016huu,Kirk:2015dpw,Liu:2015xfa,Hand:2013xua}.}
\end{figure*}

Figure~\ref{fig:Alens_comp} compares the values of $A_{\rm lens}$ and their $1\sigma$ errors
among recent cross-correlation studies between CMB lensing and cosmic shear from galaxy shapes.
The $\widehat{A}_{\rm lens}$ value obtained is slightly higher than unity but is consistent with the Planck prediction within $2 \sigma$ level.
Our result also agrees with the previous cross-correlation analyses,
although their best-fit values still have a large variation $A_{\rm lens} \simeq 0.4\text{--}1.3$ \cite[e.g.,][]{Liu:2015xfa,Harnois-Deraps:2016huu,Harnois-Deraps:2017}.
It should be noted that in the other cross-correlation studies, CMB lensing signals are dominated by those from the temperature maps,
unlike our study, in which we use the polarization map only. In addition, the redshift distributions of source galaxies are different among these measurements.

To check the robustness of our results,
Table~\ref{tab:amplitudes} shows the dependence of the amplitude on the photometric redshift estimation methods, the CMB multipoles used for the CMB lensing reconstruction, and estimators of the CMB lensing convergence. 
We also show the amplitude with respect to the \mbox{WMAP-9} cosmology. \footnote{{ We estimate $A_{\rm lens}$ for the cosmology derived from the HSC shear auto-spectrum measurement \citep{Hikage:2019}. We vary cosmological parameters within the $1\sigma$ constrains from the shear auto-spectrum, and obtain $A_{\rm lens}$ for each set of parameters. We find $\widehat{A}_{\rm lens}\simeq1.76 \text{--} 2.13$, depending on the choice of cosmological parameters. $\widehat{A}_{\rm lens}$ is consistent with $A_{\rm lens}=1$ within $2\sigma$, indicating that using the cross-spectrum does not improve the constraints from the shear auto-spectrum.}}
{ 
We find that the values of $\widehat{A}_{\rm lens}$ are all consistent with unity within $2\sigma$.
We also test statistical significance of the $\widehat{A}_{\rm lens}$ shifts by changing the analysis method, i.e., CMB multipoles and estimators. We compute the difference-amplitude, $\Delta\widehat{A}_{\rm lens}=\widehat{A}_{\rm lens}-\widehat{A}_{\rm lens}^{\rm baseline}$, for the real data and each realization of the simulation, where $\widehat{A}_{\rm lens}^{\rm baseline}$ is the value obtained from the baseline analysis. Then, we evaluate the PTEs of $\Delta\widehat{A}_{\rm lens}$, and the values of PTEs range between 0.42 and 0.88. The changes in $\widehat{A}_{\rm lens}$ compared to the baseline analysis are statistically not significant.
}

\begin{table}
\begin{center}
\caption{\label{tab:amplitudes}The amplitude of the { cross-spectrum} $\widehat{A}_{\rm lens}$ estimated with different HSC photometric redshift (photo-$z$) estimates~\citep{Tanaka:2018}, different ranges of the CMB multipoles, different CMB lensing estimators, and the different fiducial cosmology. In a fiducial case, we assume the $\Lambda$CDM cosmology with the Planck~2018\ best-fit parameters.}
\begin{tabular}{rlc}\\
\multicolumn{2}{c}{Choice of the analysis method} & $\widehat{A}_{\rm lens}$ \\\hline\hline
Photo-$z$ 
 & Ephor & $1.70 \pm 0.48$ \\ 
 & Frankenz & $1.69 \pm 0.48$ \\ 
 & MLZ & $1.83 \pm 0.51$ \\ 
 & Mizuki & $1.69 \pm 0.49$ \\ \hline
CMB multipoles
 & $\ell_{\rm max}=2500$ & $1.64\pm 0.49$ \\ 
 & $\ell_{\rm min}=700$ & $1.89\pm 0.57$ \\ \hline
CMB estimator
 & $EE$ & $1.07\pm 0.93$ \\ 
 & $EB$ & $1.65\pm 0.50$ \\\hline
 Cosmology
 & WMAP-9 & $1.99\pm 0.56$ \\ \hline\hline
\textbf{Baseline} & (Planck~2018) & $\bm{1.70\pm 0.48}$ \\
\end{tabular}
\end{center}
\end{table}

Polarized diffuse Galactic foregrounds and extra-Galactic point sources are a potential contaminant to the CMB data. The characterization of diffuse Galactic and extra-Galactic foregrounds has been derived in PB17, and here we highlight the main aspects that are relevant in our study.

The \textsc{Polarbear}{} maps have a 5$\sigma$ source detection threshold of 25~mJy. 
We mask out sources above 25~mJy to suppress contaminations from polarized 
extra-Galactic point sources. All of the sources we detect correspond to 
sources detected by either ATCA~\citep{at20g} or Planck~\citep{2016A&A...594A..26P}. 
The unmasked point sources below the 25~mJy detection threshold contribute a 
residual power, but \citet{Smith:2008:cmbpol} and 
\citet{2018ApJ...858...85P} show that this level of contribution is negligible 
in lensing auto spectra.

Polarized diffuse foregrounds are estimated based on models from the 
Planck~353~GHz and 30~GHz for dust and synchrotron, respectively 
(\textsc{Polarbear}{} collaboration et al., in prep.). PB17 fathomed the data looking 
for a signature of diffuse polarized foregrounds, found no evidence and 
obtained only upper limits. Therefore, we assume a 20\% polarization 
fraction of dust and synchrotron, which is conservative on the basis of 
all recent constraints \citep{2018arXiv180104945P,2018arXiv180706208P}. 
Moreover, we scale the modeled foregrounds to the \textsc{Polarbear}{} frequency 
assuming a modified blackbody spectral dependence for thermal dust,
with temperature $T_d\simeq 19.6\,{\rm K}$ and 
$\beta_d\simeq 1.59\pm 0.14$, and a power law for the synchrotron, 
with $\beta_s=-3.12 \pm 0.02$, consistent with most recent 
results ~\citep{2018A&A...618A.166K,2018arXiv180104945P}. 
This contamination is not found to bias lensing { auto spectrum},
indicating that  the contribution of polarized diffuse foregrounds 
is negligible in the { cross-spectrum}. We note that
varying the $\ell_{\rm min}$ in the CMB lensing reconstruction does not 
significantly change the result~(Table~\ref{tab:amplitudes}). 
This supports the foreground contribution as minor in
our results, as diffuse foregrounds have larger contributions 
in low $\ell$ regions.

Both CMB lensing and cosmic shear have contributions from the nonlinear evolution
of the large-scale structure and post-Born corrections (e.g., \citealt{Cooray:2002mj,Takada:2004,KrauseHirata:2010,Namikawa:2016b,Pratten:2016,Fabbian:2017wfp}). Consequently, the nonlinear evolution of the gravitational potential (or density perturbations) and the post-Born corrections lead to additional contributions in the { cross-spectrum}. However, its contribution is known to be below $1\%$, and is negligible at the current level of sensitivity \citep{Boehm:2016,Merkel:2017,Boehm:2018,Beck:2018}.

The intrinsic alignment produces the cross-correlation between CMB lensing and cosmic shear (e.g., \citealt{Hirata:2004rp}). However, \citet{Hikage:2019} shows, using the cosmic shear { auto power spectrum}, that the amplitude of the intrinsic alignment is consistent with zero, implying that the intrinsic alignment is also not significant in our shear data as compared to the statistical uncertainty. { As shown in Figure~10 in \citet{Hikage:2019}, the observed amplitude of intrinsic alignment $A_{\rm IA}$ is consistent with a red galaxy only model in which only red galaxies are assumed to have intrinsic alignments. This model predicts $A_{\rm IA}\lesssim 2$ within the redshift range where the cosmic shear measurement was performed. According to \cite{Hall:2014nja} and \cite{Chisari:2015}, this size of intrinsic alignment yields about $5\text{--}10\%$ contamination to the cross-correlation signal, which is much smaller than the statistical uncertainty in this measurement \citep[also see][]{Troxel:2014kza,Larsen:2015aoa}.}

\section{Conclusion} \label{sec:conclusion} 
We have presented a new measurement of the { cross-spectrum} between the CMB lensing map from the \textsc{Polarbear}{} experiment and the cosmic shear field from the Subaru HSC survey.
We measured a gravitational lensing amplitude of $\widehat{A}_{\rm lens}=1.70\pm 0.48$, with respect to the Planck $\Lambda$CDM cosmology, which represents the detection of a non-zero cross-correlation at 3.5$\sigma${} significance.
Although there have been several significant detections of such cross-spectra
during the past several years~\citep[e.g.,][]{Hand:2013xua,Kirk:2015dpw,Liu:2015xfa,Singh:2016xey,Harnois-Deraps:2016huu,Omori:2018},
in this paper we presented the first detection of the cross-correlation between CMB lensing
and cosmic shear from galaxy shapes, solely from the CMB {\it polarization} map, i.e., without relying on the CMB temperature measurement.
Both the high galaxy number density of $n_g=23\,{\rm arcmin^{-2}}${} for HSC and the deep CMB map of ${\sim}$6\,$\mathrm{\mu K}$-arcmin\ for \textsc{Polarbear}{} lead to this measurement of the { cross-spectrum},
even for a relatively small overlapping area of ${\sim} 11$~deg$^2$.
We also note that this work represents the first cross-correlation measurement between the HSC cosmic shear and CMB lensing.

Both CMB and cosmic shear measurements directly trace the mass distribution in the universe through gravitational lensing.
The cross-correlation analysis of these two types of datasets is robust against instrumental and astronomical systematics
{ that are additive, since the bias in the two data sets are unlikely to be correlated.
This in turn can constrain possible multiplicative bias in the weak-lensing dataset, validating the calibration for measurements of the mass distribution.}
The cross-correlation is sensitive to the mass distribution in the medium redshift range of $z \sim 1$,
and is complementary to auto spectra of CMB lensing and cosmic shear.
Significant improvements in the measurement of the cross-correlation, which is expected in the next decade,
will contribute to better understanding of a neutrino mass, dark energy, and its possible time evolution.

The lensing maps from the CMB polarization, in contrast to those from the CMB temperature, are less contaminated by Galactic or extra-Galactic foregrounds, and will become more accurate than the temperature lensing maps in future deep surveys.  Even though our analysis is based on a \textsc{Polarbear}{} field covering only several square degrees in area, the depth of the map is comparable to what we expect to achieve in future experiments, such as at Simons Observatory~\citep{SimonsObservatory}.\footnote{\url{https://simonsobservatory.org/}} Similarly, the Subaru HSC cosmic shear map is one of the deepest maps to date, and can be seen as a precursor of the Large Synoptic Survey Telescope~(LSST).\footnote{\url{https://lsst.slac.stanford.edu/}} Wide-field space-based telescopes such as WFIRST and Euclid are planned to be launched in the 2020s and will provide deep, dense, and highly-resolved galaxy images, with the galaxy number density comparable to or better than that of the HSC survey. These future datasets could provide cosmological measurements at a sub-percent accuracy.
The shear calibration requirement of LSST sets a concrete goal for the future dataset to achieve ${\sim}0.5\%$ accuracy of the cross-correlation between CMB lensing maps and galaxy cosmic shear maps~\citep{Schaan:2016ois,S4ScienceBook,SimonsObservatory}.
While CMB temperature data suffer from foreground contaminations, CMB polarization measurements provide a better path to achieve this goal~\citep{Schaan:2016ois}. Our results serve as a step forward to future experiments. For instance, we performed a detailed study on possible systematic errors and found no significant bias, placing an upper limit on ${\sim} 1$\% level in the lensing amplitude measurement.  These systematic estimates are primarily limited by statistical uncertainty in our systematics-error study, while systematic errors are likely to be further reduced in future datasets. Therefore, our work demonstrates the potential and promise of this cross-correlation methodology to provide insight into fundamental problems of cosmology, such as the nature of neutrinos and dark energy.

\acknowledgments
We are thankful for fruitful discussions and comments provided by Simone Ferraro and Emmanuel Schaan.

The Hyper Suprime-Cam (HSC) collaboration includes the astronomical communities of Japan and Taiwan, and Princeton University.  The HSC instrumentation and software were developed by the National Astronomical Observatory of Japan (NAOJ), the Kavli Institute for the Physics and Mathematics of the Universe (Kavli IPMU), the University of Tokyo, the High Energy Accelerator Research Organization (KEK), the Academia Sinica Institute for Astronomy and Astrophysics in Taiwan (ASIAA), and Princeton University.  Funding was contributed by the FIRST program from Japanese Cabinet Office, the Ministry of Education, Culture, Sports, Science and Technology (MEXT), the Japan Society for the Promotion of Science (JSPS), Japan Science and Technology Agency (JST), the Toray Science  Foundation, NAOJ, Kavli IPMU, KEK, ASIAA,  and Princeton University.

The Pan-STARRS1 Surveys (PS1), which are used for the photometry and astrometry calibration, have been made possible through contributions of the Institute for Astronomy, the University of Hawaii, the Pan-STARRS Project Office, the Max-Planck Society and its participating institutes, the Max Planck Institute for Astronomy, Heidelberg and the Max Planck Institute for Extraterrestrial Physics, Garching, The Johns Hopkins University, Durham University, the University of Edinburgh, Queen's University Belfast, the Harvard-Smithsonian Center for Astrophysics, the Las Cumbres Observatory Global Telescope Network Incorporated, the National Central University of Taiwan, the Space Telescope Science Institute, the National Aeronautics and Space Administration under Grant No. NNX08AR22G issued through the Planetary Science Division of the NASA Science Mission Directorate, the National Science Foundation under Grant No. AST-1238877, the University of Maryland, and Eotvos Lorand University (ELTE).
 
This paper makes use of software developed for the Large Synoptic Survey Telescope. We thank the LSST Project for making their code available as free software at \url{http://dm.lsst.org}.

This work is based in part on data collected at the Subaru Telescope and retrieved from the HSC data archive system, which is operated by the Subaru Telescope and Astronomy Data Center at National Astronomical Observatory of Japan.

The \textsc{Polarbear}{} project is funded by the National Science Foundation under Grants No. AST-0618398 and No. AST-1212230.
The James Ax Observatory operates in the Parque Astron\'omico Atacama in Northern Chile under the auspices of the Comisi\'on Nacional de Investigaci\'on Cient\'ifica y Tecnol\'ogica de Chile (CONICYT).
The James Ax Observatory would not be possible without the support of CONICYT in Chile.
All silicon wafer-based technology for \textsc{Polarbear}{} was fabricated at the UCB Nanolab.
This research used resources of the Central Computing System, owned and operated by the Computing Research Center at KEK, the HPCI system (Project ID:hp150132), and the National Energy Research Scientific Computing Center~(NERSC), a DOE Office of Science User Facility supported by the Office of Science of the U.S. Department of Energy under Contract No. DE-AC02-05CH11231.
This work was supported in part by World Premier International Research Center Initiative (WPI Initiative), MEXT, Japan, and MEXT Grant-in-Aid for Scientific Research on Innovative Areas (JP15H05887, JP15H05891, JP15H05892, JP15H05893).
In Japan, this work was supported by MEXT KAKENHI Grant Number 21111002 and JP18H05539, JSPS KAKENHI grant Nos. JP26220709, JP24111715, JP26800125, and the JSPS Core-toCore Program.
In Italy, this work was supported by the RADIOFOREGROUNDS grant of the European Union's Horizon 2020 research
and innovation programme (COMPET-05-2015, grant agreement number 687312) as well as by the INDARK INFN Initiative
and the COSMOS network of the Italian Space Agency~(cosmosnet.it).
Support from the Ax Center for Experimental Cosmology is gratefully acknowledged.

The KEK authors acknowledge support from KEK Cryogenics Science Center.
The APC group acknowledges travel support from Labex UNIVEARTHS.
The Melbourne group acknowledges support from the University of Melbourne and an Australian Research Council’s Future Fellowship (FT150100074).
YC acknowledges the support from the JSPS KAKENHI Grant Number 18K13558 and 18H04347.
TN acknowledges the support from the Ministry of Science and Technology (MOST), Taiwan, R.O.C. 
through the MOST research project grants (no. 107-2112-M-002-002-MY3). 
HM acknowledges the support from JSPS KAKENHI Grant Number JP18H04350.
MO acknowledges the support from JSPS KAKENHI Grant Number JP18K03693.
RT acknowledges the support from Grant-in-Aid for Scientific Research from the JSPS Promotion of Science (17H01131).
Numerical computations were in part carried out on Cray XC50 at Center for Computational Astrophysics, National Astronomical Observatory of Japan.
AK acknowledges the support by JSPS Leading Initiative for Excellent Young Researchers (LEADER) and by the JSPS KAKENHI Grant Number JP16K21744.
MA acknowledges support from CONICYT UC Berkeley-Chile Seed Grant (CLAS fund) Number 77047, Fondecyt project 1130777 and 1171811, DFI postgraduate scholarship program and DFI Postgraduate Competitive Fund for Support in the Attendance to Scientific Events.
GF acknowledges support from the European Research Council under the European Union's Seventh Framework Programme (FP/2007-2013) / ERC Grant Agreement No. [616170].
FM acknowledges the support by the JSPS { F}ellowship (Grant number JP17F17025).
JSS is supported by the NSF Graduate Research Fellowship.

\bibliographystyle{aasjournal}

\end{document}